\documentclass[prl,twocolumn,superscriptaddress,showpacs,floatfix,amsfonts]{revtex4}
\usepackage{graphicx}% Include figure files
\usepackage{bm}
\usepackage{amsmath}
\usepackage{amssymb}
\begin{document}
\preprint{}
\title{High Angular Momentum Core States and Impurity Effects in
the Mixed State of $d$-Wave Superconductors}
\author{Jian-Xin Zhu}
\affiliation{Theoretical Division, MS B262, 
Los Alamos National Laboratory, Los Alamos, 
New Mexico 87545}
\author{C. S. Ting}
\affiliation{Texas Center for Superconductivity, University of Houston, Houston,
Texas 77204
}
\author{A. V. Balatsky}
\affiliation{Theoretical Division, MS B262,
Los Alamos National Laboratory, Los Alamos, 
New Mexico 87545}
%\date{\today}

\begin{abstract}
{ 
The local quasiparticle spectra around 
the vortex core of a $d$-wave superconductor in the mixed state 
is studied by solving  the lattice Bogoliubov-de Gennes equations
self-consistently. It is shown that in addition to the zero-energy  
states, there also exist core states of high angular momentum. 
A nonmagnetic unitary impurity sitting at the core center is nondestructive
to these core states and the zero-energy resonant state induced by itself 
is still visible in the local quasiparticle spectrum.
The calculated imaging shows a fourfold ``star'' shape of the 
local density of states whose orientation is energy dependent.
It is also found that, although the zero-energy states are extended, 
the core states of finite energy level could be localized.
}
\end{abstract} 
\maketitle

In the Abrikosov vortex state (also called mixed state or Schubnikov phase) 
of type-II superconductors, the magnetic flux penetrates into the 
system in quantized units $\Phi_{0}=hc/2e$. In the region of vortex core, the 
superconducting order parameter (or pair potential) 
is considerably suppressed from its zero-field 
value $\Delta_0$, going to zero at the vortex core center at a distance of 
several coherence lengths $\xi_0$. This reduction of the order 
parameter can affect various physical properties of the superconductor, for 
example, the nature of local electronic states around the Fermi energy. 
The vortex core problem of an $s$-wave superconductor was first studied 
in the classic papers of Caroli, de Gennes, and Matricon~\cite{CdeGM64}.
By solving approximately the Bogoliubov-de Gennes equations for an isolated 
vortex line in the limit of $\kappa=\lambda/\xi_0 \gg 1$, 
where $\lambda$ is the 
magnetic penetration depth, 
these authors showed the existence of the low-lying bound
quasiparticle states inside an $s$-wave vortex core. Later on, Bardeen {\em 
et al.}~\cite{BKJT69} extended the calculation to all values of $\kappa$ 
by determining the pairing potential and magnetic field with a variational 
expression for the free energy, and the qualitative conclusion is similar to
that of Caroli {\em et al.}. Recent 
theoretical 
attempts~\cite{SHDS89,Klein89,Gygi90,UDB90,PK93,ZZS95} 
at understanding
the electronic states of vortex lines in type-II superconductors 
was revived by STM experiments on NbSe$_2$~\cite{HRDVW89}, 
which not only showed a strong enhancement of the zero-bias
tunneling conductance as a manifestation of the presence of
bound states in the vortex core, but also provided images of vortex.
For an $s$-wave superconductor, the energy gap open at the Fermi  surface 
is a constant. One can naively think that, 
the spatial variation  of the pair potential is analogous to 
a potential well for the quasiparticle, 
with depth $\Delta_0$ and radius $\xi_0$. The energy levels of the quasiparticle
bound states is roughly the multiples of $E_0=\hbar^2/2m\xi_0^2\approx 
\Delta_0^2/E_F$, where the energy is measured relative to the Fermi energy 
$E_F$. For a $d$-wave superconductor 
as recently established in CuO$_2$-based high-$T_c$ cuprates, the energy
gap has a maximum along the copper oxide bonds and 
is closed along the diagonals of square CuO$_2$ lattice.
In some respect, the spatial variation of the pair potential 
is analogous to a potential well carved with four notches along the nodal
directions, so that
quasiparticles can leak through the notches. It was shown for the first time 
by  Wang and MacDonald~\cite{WM95} that 
there appears a single broad peak at zero energy
in the local density of states (LDOS) at the center of a pure
$d$-wave vortex core so that the quasiparticle states are virtually bound.  
Recently, the split-peak structure 
around zero bias in the local
differential tunneling conductance at the vortex core center in 
YBa$_2$Cu$_3$O$_{7-\delta}$~\cite{MREWF95} 
and Bi$_2$Sr$_2$CaCu$_2$O$_{8+\delta}$~\cite{PHGN00,HKRL01}
has been 
observed by scanning tunneling microscopy (STM).
These experimental observations stimulated intensified study  
of the quasiparticles in the vortex core of high-$T_c$ 
cuprates~\cite{HOTK97,MKM97A,ABKZ97,FT98,TIK99,YK99,HL00,ABH00,WXS00,KLW01,Zhu01c,BG01},
leading to a variety of scenarios for the explanation of experimental data.
No consensus on the mechanism has emerged. However, on the other hand, 
additional features of quasiparticle states of a pure $d$-wave vortex 
have not been well addressed even in the framework of BCS-type description.
Specifically, nearly all of the theoretical and experimental works on the 
quasiparticle states of $d$-wave
vortices focus on the LDOS directly on the core center, 
whether there exists high angular momentum core states 
have not been paid much attention. 
Furthermore, the imaging of the tunneling conductance
of a $d$-wave vortex, as in the impurity case~\cite{PHLE00}, 
has not been experimentally
available. In this paper, we present a complete study of 
the electronic structure 
around one of the $d$-wave vortices forming the Abrikosov square lattice
by solving self-consistently the Bogoliubov-de Gennes (BdG) equations within a 
tight-binding model, and new features are exposed 
for the first time.
%It is shown that the LDOS around the vortex 
%exhibits the split peaks with the spacing increasing with 
%the distance away from the core center. These peaks are mainly contributed 
%from the tunneling into the high angular momentum core states. 
 
The electronic structure of a $d$-wave superconductor can be described by the
quasiparticle wavefunctions 
$
\left( \begin{array}{c} u_{i}^{n} \\
v_{i}^{n}\end{array} \right)
$
which satisfy the BdG equations
\begin{equation}
\sum_{j} \left(
\begin{array}{cc}
{\mathcal H}_{ij} & \Delta_{ij}  \\
\Delta_{ij}^{*} & -{\mathcal H}_{ij}^{*}
\end{array}
\right) \left(
\begin{array}{c}
u_{j}^{n} \\ v_{j}^{n}
\end{array}
\right)
=E_{n}
\left(
\begin{array}{c}
u_{i}^{n} \\ v_{i}^{n}
\end{array}
\right)  \;.
\label{EQ:BdG}
\end{equation}
Here  the summation is over the nearest neighbor sites. 
The single particle Hamiltonian reads
${\mathcal H}_{ij}=-t e^{i\varphi_{ij}} \delta_{
i+\tau,j} +
(U_{i}-\mu)\delta_{ij}$ 
where $t$ is the hopping integral, $\mu$ is the chemical potential,
$U_{i}$ is the single particle 
potential describing, if any, the effects of impurities, defects, 
or crystal field, and
$\tau=\pm\hat{e}_{x},\pm\hat{e}_{y}$ 
are the unit vectors along the crystalline $x$ 
and $y$ axes, respectively.  
In the mixed state, the magnetic field effect
is included through the
Peierls phase factor $\varphi_{ij}=\frac{\pi}{\Phi_{0}}
\int_{\mathbf{r}_{j}}^{\mathbf{r}_{i}} 
\mathbf{A}(\mathbf{r})\cdot d\mathbf{r}$.
By assuming the superconductor under consideration is in the extreme type-II
limit where $\kappa$ goes to
infinity so that the screening effect from the supercurrent is negligible. 
The vector potential $\mathbf{A}$ can then be approximated by the
solution $\nabla \times \mathbf{A}=H \hat{\mathbf{z}}$ 
where $H$ is the magnetic
field externally applied along the $c$ axis. The enclosed flux density
within each plaquette is given by $\sum_{\square} \varphi_{ij}
=\frac{\pi Ha^{2}}{\Phi_0}$. 
Notice that the quasiparticle
energy is measured with respect to the Fermi energy.
$\Delta_{ij}$ is the spin-singlet $d$-wave bond pair potential, which
 is
subject to the self-consistency condition:
\begin{equation}
\Delta_{ij}=\frac{V}{2}\sum_{n,E_n>0}
(u_{i}^{n}v_{j}^{n*}
+v_{i}^{n*}u_{j}^{n}
)
\tanh \left( \frac{E_{n}}{2k_{B}T}\right)\;,
\end{equation}
with $V$ the strength of
nearest neighbor effective attraction between electrons.
Hereafter we measure the length in units of the
lattice constant $a$ and the energy in units of the hopping integral $t$.
Within the Landau gauge the vector potential can be written as
$\mathbf{A}=(-H y,0,0)$ where $y$ is the $y$-component of the position
vector $\mathbf{r}$.
We introduce the magnetic translation operator
$\mathcal{T}_{mn}\mathbf{r}=\mathbf{r}+\mathbf{R}$
where the translation vector $\mathbf{R}=m N_x \hat{\mathbf{e}}_{x} +n N_y
\hat{\mathbf{e}}_{y}$ with $N_{x}$ and $N_{y}$ the linear dimension of the unit 
cell of the vortex lattice.
To ensure different $\mathcal{T}_{mn}$ to be commutable with each other,
we have to take the strength of magnetic field so that the flux enclosed
by each unit cell has a single-particle flux quantum, i.e, $2\Phi_0$.
Therefore, the translation property of the superconducting order parameter
is $\Delta (\mathcal{T}_{mn}\mathbf{r})=e^{i\chi(\mathbf{r},\mathbf{R})}
 \Delta(\mathbf{r})$ where the phase accumulated 
 by the order parameter upon the 
translation   is
$\chi(\mathbf{r},\mathbf{R})=\frac{2\pi}{\Phi_0}\mathbf{A}(\mathbf{R})
 \cdot \mathbf{r}-4mn\pi$. 
 From this property, we can obtain the magnetic Bloch theorem for
the wavefunction of the BdG equations:
\begin{equation}
\left(
\begin{array}{c}
u_{\mathbf{k}}(\mathcal{T}_{mn}\tilde{\mathbf{r}}) \\
v_{\mathbf{k}}(\mathcal{T}_{mn}\tilde{\mathbf{r}})
\end{array}
\right)
 = e^{i\mathbf{k}\cdot \mathbf{R}}
\left(
\begin{array}{c}
e^{i\chi(\mathbf{r},\mathbf{R})/2} u_{\mathbf{k}}(\tilde{\mathbf{r}}) \\
e^{-i\chi(\mathbf{r},\mathbf{R})/2}v_{\mathbf{k}}(\tilde{\mathbf{r}})
\end{array}
\right) \;.
\end{equation}
Here $\tilde{\mathbf{r}}$ is the position vector defined within a given unit
cell and $\mathbf{k}=\frac{2\pi l_x}{M_x N_x}\hat{\mathbf{e}}_{x} +
\frac{2\pi l_y}{M_y N_y}\hat{\mathbf{e}}_{y}$ with 
$l_{x,y}=0,1,\dots,M_{x,y}-1$
are the wave vectors defined in the first Brillouin zone of the vortex 
lattice and $M_x N_x$ and  $M_y N_y$ are  the linear dimension of the whole 
system. We use exact diagonalization method to
solve the BdG equation~(\ref{EQ:BdG}) self-consistently.
The thermally broadened local density of states is then evaluated according to   
\begin{equation}
\rho_{i}(E)=-\frac{2}{N_c} \sum_{\mathbf{k},n} 
[\vert u^{n}_{\mathbf{k},i} \vert^{2} f^{\prime}(E^{n}_{\mathbf{k}}-E)
+\vert v^{n}_{\mathbf{k},i}\vert^{2} f^{\prime}(E^{n}_{\mathbf{k}}+E)]
\;,
\end{equation}  
where a prefactor comes from the spin degeneracy, and 
$f^{\prime}(E)$ is the derivative of the Fermi distribution function 
$f(E)=1/[\exp(E/k_{B}T)+1]$ with respect to the energy $E$,  
$N_{c}=M_x \times M_y$ is the number of magnetic unit cells.
$\rho_{\mathbf{i}}(E)$ is proportional to the local differential tunneling
conductance at low temperatures which could be   
measured by STM experiments~\cite{Tinkham96}. 

\begin{figure}
\includegraphics{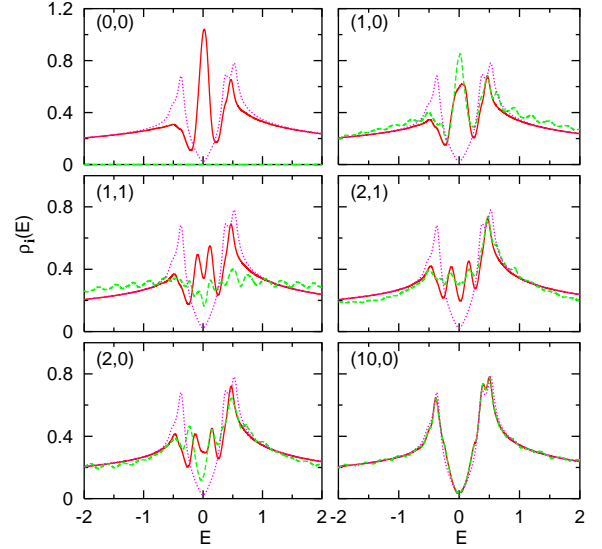}% Here is how to import EPS art
\caption[*]{
The local density of states as a function of energy at various distance
away from the $d$-wave vortex core center $(0,0)$. The coordinate is 
measured in units of the lattice constant. The solid line in each panel
shows the result in the clean limit. The result with a nonmagnetic 
unitary impurity sitting at the core center are shown by the dashed line.
The bulk density of states in the absence of the magnetic 
field and impurities is also plotted for a reference by the dotted line.
For the chosen unit cell size $42 \times 21$,  the site (10,0) is the
midpoint between two nearest neighboring 
vortices along the $x$ direction. 
The other parameter values: $V=1.2$, $n_f=0.85$, and $T=0.01$.    
}
\label{FIG:LDOS}
\end{figure}

In the following calculation, we take $V=1.2$, 
and the filling factor $0.85$ 
by adjusting the chemical potential. The temperature is chosen 
to be $T=0.01$. 
Without loss of generality, 
we consider the 
unit cell of size $N_x\times N_y= 42\times 21$,
and the number of the unit cells
$N_c=21\times 42$. This choice will give us a square vortex 
lattice. The numerical calculation shows as expected that 
each unit cell accommodates two superconducting vortices each carrying a 
flux quantum $\Phi_{0}$.  The spatial variation of the $d$-wave order 
parameter is  similar to Fig.~1 of Ref.~\cite{WM95}: It decreases continuously
to zero from the its bulk value as the vortex core center is approached in the 
length scale $\xi_0=\hbar v_{F}/2 \Delta_{max}\approx 5$, 
with the depleted region
extending farther in the diagonal direction of the square lattice. Here 
$\Delta_{max}\approx 0.38$ 
is the gap edge defined as the energy position 
at which the bulk DOS becomes maxima, and $v_{F}$ is the Fermi velocity.   
For the given model parameter, the induced extended $s$-wave order parameter
around the vortex is so small that its effect on the LDOS should be negligible.
In Fig.~\ref{FIG:LDOS} we plot the LDOS as a function
of energy at various distance from the vortex core in a $d$-wave 
superconductor. The solid line in each panel represents the result 
of a clean system, while the dashed line for the case that a unitary
impurity ($U_0=100$) sits at the vortex core center.
For comparison, we have also displayed the bulk LDOS (the dotted lines) in the 
absence of magnetic field and impurities. We choose
the core center as the coordinate origin (0,0). For the given unit cell size,
the site (10,0) is the midpoint of two nearest neighboring vortices along the
$x$ direction. Notice that the three curves in the right-bottom panel 
almost coincide with each other, indicating that the LDOS spectrum
at the midpoint has recovered the bulk DOS.
The asymmetry of the LDOS spectrum with respect to the 
Fermi energy comes from the breaking of the particle-hole symmetry.
In the absence of the unitary impurity,  
the LDOS at the core center $i=(0,0)$ 
shows a single resonant peak around 
the Fermi energy. Although the spectrum at the center does not exhibit a
split-peak structure, the suppression of the coherent peak at the gap edge  
$\Delta_{max}$ seems to be consistent with the experimental observations. 
At this point, we also calculate the LDOS at the core center with 
various values of the bulk order parameter and find that the broad zero-energy
peak always shows up, which leads us to conclude that the quasiparticle 
excitations along the nodal direction on the Fermi surface should be the origin. 
At a finite distance away from the core center, the single peak in the LDOS
at the core center continuously evolves into two peaks, which are located at 
$\pm \delta E(i)$. The spacing between these two 
peaks, i.e., $2\delta E(i)$  
increases with the distance away from the core center. 
In addition, the amplitude of these peaks  
decreases with the distance from the core center and finally they merge 
into the continuum of scattering state with $E>E_{max}$. 
We remark that, since the induced extended $s$-wave order parameter is 
almost negligible, this origin for the obtained split peaks should be
excluded. 
Considering the distance dependence of these subgap peaks in the LDOS, 
we believe instead that they result from the core states with a 
high angular momentum. The effect of a nonmagnetic impurity in the unitary 
limit is twofold: First, it plays the role of a pinning center. 
We introduce such an impurity into the unit cell randomly.  
It is found that when this impurity has a distance from the vortex 
core center within the range of a coherence length, the vortex will 
be dragged onto the impurity site. The order parameter profile 
of the pinned vortex is quite similar to that of a clean case. 
Out of this range, the pinning effect is rather weak. 
Second, it is both theoretically~\cite{SBS96} and 
experimentally~\cite{PHLE00,HPGN99,YHLK99} 
well established
that a nonmagnetic impurity itself will
induce resonant states in a $d$-wave superconductor 
in contrast to its counterpart in a conventional $s$-wave superconductor.
For the convenience of discussion, we consider the lattice as a bipartite, 
i.e, those sites $i=(ix,iy)$ with index having {\em even}
$(-1)^{ix+iy}$ form sublattice $A$, which contains the core center,
while those with index having
{\em odd} $(-1)^{ix+iy}$ form sublattice $B$.
The panels in the left column of Fig.~\ref{FIG:LDOS} display the LDOS 
at the sites belonging to $A$, and the panels in the right column 
show the LDOS at the sites belong to $B$. 
When such an impurity pins the vortex, 
the LDOS spectrum at the core center vanishes due to the strong potential
scattering from the impurity [See the dashed line as shown in 
the panel of Fig.~\ref{FIG:LDOS} corresponding 
to the site (0,0)]. However, the intensity of the single zero-energy peak 
in the LDOS at the nearest neighbor site of the core center is strongly
enhanced. This enhancement comes from the additional contribution of the 
near-zero-energy resonant state induced directly by the impurity 
itself~\cite{SBS96}. 
As the measured point is away from the core center, the LDOS exhibits no 
central peak (near the Fermi energy) at the sites belonging to the sublattice 
$A$. In particular, the density of states at zero energy is more depressed 
on these sites as compared to the impurity free case. 
Nevertheless, the LDOS on the sites belonging to the sublattice $B$ has a 
zero-energy peak, the intensity of which decays with the increased distance
away from the core center. This pattern of the LDOS around the zero energy 
is solely caused by the impurity-induced resonant state, which 
has an oscillation behavior~\cite{SBS96,ZTH00}.
 Also interestingly, it can be seen that the location
of the LDOS peaks associated with the high angular momentum 
core states is quite insensitive to the impurity scattering. In this sense,
we can say  
that the unitary impurity  is nondestructive to the vortex from the point of
view of core states. The above effect of the single nonmagnetic unitary 
impurity on the electronic structure of a $d$-wave vortex is dramatically 
different from its effect in the case of an $s$-wave vortex. Since there
is no resonant states induced around this impurity in an $s$-wave 
superconductor, it drives the energy position of the 
low-level core states away from the Fermi energy, and simultaneously the 
intensity of the corresponding peaks in the LDOS becomes weaker with the 
impurity strength~\cite{Han00}.

To take a closer inspection of  the nature of the electronic structure
around a $d$-wave vortex, we calculate the spatial distribution 
of the LDOS around the
vortex core at the fixed energy $E_1=0$ and $E_{2}=0.15$. $E_1$ is the peak
position of the LDOS at the core center, $E_{2}$ is that 
at the site (2,1). For the clean case, we find strong anisotropy of the spatial
distribution of the LDOS, which has a fourfold ``star'' shape. 
At zero energy ($E_1=0$), the magnitude of the LDOS 
has tails extending along the diagonals of the  
square lattice, and decays rapidly along the CuO$_2$ bonds.
This behavior can be understood as follows: 
When the energy gap vanishes along the nodal directions, the quasiclassical 
orbital of these low-lying quasiparticle states is not closed. Therefore, 
the zero-energy states are extended along these directions and consequently, 
in contrast to the $s$-wave case, they are not truly bound core states.
However, at a finite energy ($E_2=0.15$), the situation is reversed and the LDOS
decays more slowly along the CuO$_2$ bonds than along the diagonals of the square
lattice, that is, the ``star'' shape is rotated by almost $\pi/4$.
In particular, the decay even along the bonds is so short ranged that the core
states at this energy is approximately localized.  
This energy-dependent orientation of the ``star'' shape 
around a $d$-wave vortex arises from the $d$-wave pairing symmetry itself,
which is in striking contrast to the $s$-wave case, the shape rotation 
was explained by introducing a perturbation term~\cite{Gygi90}.
When the impurity is added on the core center, the above conclusion 
is not changed except that the LDOS directly 
at the impurity site and the ``star'' becomes more extended.

To summarize, we have presented a study of the quasiparticle states 
around the vortex core of a $d$-wave superconductor in the mixed state. 
We have shown that, in addition to the zero-energy core states, there also 
exist core states with high angular momentum. It has also been found that 
a nonmagnetic impurity is not destructive to these core states, and the 
impurity-induced near-zero energy resonant state~\cite{SBS96}
 can still be visible 
in the LDOS. The calculated images demonstrate that all core states are not 
truly localized. So far, it seems that all theoretical and 
experimental studies focused on the local electronic structure 
at the vortex core center, on which much controversy just arises. 
The test of the existence of the high angular momentum core states, 
as well as the effects of the single atomic impurity, 
is now accessible to the experiments. 
Currently, the STM is the technique of choice that allows the observation
of vortex states. It 
has recently been improved in both the spatial (atomic scale) 
and energy (sub-meV) resolution~\cite{PHLE00}. 
We suggest that the STM tip be scanned horizontally in the region 
of several lattice constants around the vortex core center
to observe the predicted changes in intensity and energy position of peaks. 
If the above predictions are indeed experimentally observed,
it will help us to understand the delicate nature of the electronic structure 
of high-$T_c$ vortices.

\begin{figure}
\includegraphics{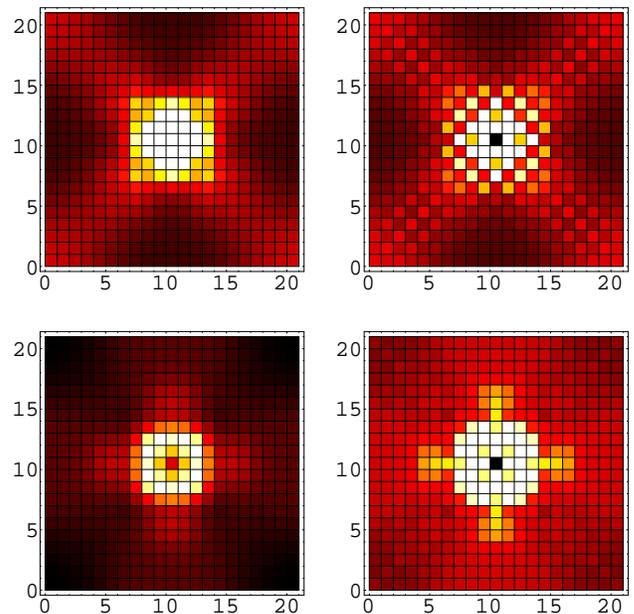}
\caption[*]{The spatial distribution of the LDOS 
around a $d$-wave vortex at the energy $E_{1}=0$ (top row) and $E_{2}=0.15$ 
(bottom row). The left column shows the results in the clean case. 
Those with a nonmagnetic unitary impurity sitting at the core center are 
displayed on the right column.
The measured size is $21\times 21$. 
The other parameter values are the same 
as Fig.~\ref{FIG:LDOS}.
}
\label{FIG:IMAGE}
\end{figure}

{\bf Acknowledgments}: We wish to thank A.H. MacDonald and K.K. Loh
for useful discussions. 
This work was supported by the Department of Energy through the Los Alamos 
National Laboratory and 
by the Texas Center for Superconductivity at the University of 
Houston.

\end{document}